%Paper: hep-th/9211138
%From: Jean-Bernard Zuber <zuber@amoco.saclay.cea.fr>
%Date: 30 Nov 92 16:16:33+0100

%%%%%%%%%%%%%%%%%%%%%%%%%%%%%%%%%%%%%%%%%%%%%%%%%%%%%%%%%%%%%%%%%%%%%%%%%%
%                        FUSION POTENTIALS I                             %
%  									 %
%                by P. Di Francesco, and J.-B. Zuber    		 %
%                TEX file, using harvmac.tex macros  		         %
%									 %
%                    SPhT  92/138  hep-th/9211138     			 %
%%%%%%%%%%%%%%%%%%%%%%%%%%%%%%%%%%%%%%%%%%%%%%%%%%%%%%%%%%%%%%%%%%%%%%%%%%
\input harvmac
\overfullrule=0mm

%Macros
%%%%%%%%%%%%%%%%%%%%%%%%%%%%%%%%%%%%%%%%%%%%%%%%%%%%%%%%%%%%%%%
%%%%%%%%%%%%%%%%%%%DEFINITIONS%%%%%%%%%%%%%%%%%%%%%%%%%%%%%%%%%
%
\def\frac#1#2{\scriptstyle{#1 \over #2}}

%
%%%%%%%%%%%%%%%%%%%CALIGRAPHIC LETTERS%%%%%%%%%%%%%%%%%%%%%%%%%
%

\def\CM{{\cal M}}		
\def\CP{{\cal P}}	\def\CQ{{\cal Q}}	\def\CR{{\cal R}}
		\def\CU{{\cal U}}
\def\CV{{\cal V}}		

\def\({ \left( }
\def\){ \right) }
%%%%%%%%%%%%%%%%%%%%MATH CHARACTERS%%%%%%%%%%%%%%%%%%%%%%%%%%%%
%

%\font\numbers=cmu10 scaled\magstep1

\def\IR{\relax{\rm I\kern-.18em R}}
\font\cmss=cmss10 \font\cmsss=cmss10 at 7pt
\def\IZ{\relax\ifmmode\mathchoice
{\hbox{\cmss Z\kern-.4em Z}}{\hbox{\cmss Z\kern-.4em Z}}
{\lower.9pt\hbox{\cmsss Z\kern-.4em Z}}
{\lower1.2pt\hbox{\cmsss Z\kern-.4em Z}}\else{\cmss Z\kern-.4em Z}\fi}
\def\inbar{\,\vrule height1.5ex width.4pt depth0pt}
\def\IB{\relax{\rm I\kern-.18em B}}
\def\IC{\relax\hbox{$\inbar\kern-.3em{\rm C}$}}
\def\ID{\relax{\rm I\kern-.18em D}}
\def\IE{\relax{\rm I\kern-.18em E}}
\def\IF{\relax{\rm I\kern-.18em F}}
\def\IG{\relax\hbox{$\inbar\kern-.3em{\rm G}$}}
\def\IH{\relax{\rm I\kern-.18em H}}
\def\II{\relax{\rm I\kern-.18em I}}
\def\IK{\relax{\rm I\kern-.18em K}}
\def\IL{\relax{\rm I\kern-.18em L}}
\def\IM{\relax{\rm I\kern-.18em M}}
\def\IN{\relax{\rm I\kern-.18em N}}
\def\IO{\relax\hbox{$\inbar\kern-.3em{\rm O}$}}
\def\IP{\relax{\rm I\kern-.18em P}}
\def\IQ{\relax\hbox{$\inbar\kern-.3em{\rm Q}$}}
\def\IGa{\relax\hbox{${\rm I}\kern-.18em\Gamma$}}
\def\IPi{\relax\hbox{${\rm I}\kern-.18em\Pi$}}
\def\ITh{\relax\hbox{$\inbar\kern-.3em\Theta$}}
\def\IOm{\relax\hbox{$\inbar\kern-3.00pt\Omega$}}

%%%%%%%%%%%%%%%%%%%%%%%%%%%%%%%%%%%%%%%%%%%%%%%%%%%%%%%%%%%%%%%%%

%Mes Macros

\def\oh{{1\over 2}}\def\un{{\bf 1}}
%\def\Z{_N Z}

%%%%%%%%%%%%%%%%%%%%Greek letters%%%%%%%%%%%%%%%%%%%%%%%%%%%%%%%%%%
\def\Ga{\alpha}\def\Gb{\beta}\def\Gc{\gamma}
\def\Ge{\epsilon}
\def\Gth{\theta}
\def\Gl{\lambda}

\def\Gs{\sigma}

%%%%%%%%%%%%%%%%

\def\mod{{\rm mod\,}}

%nouvelles macros de ce papier
\def\rep{representation} \def\Che{Chebishev\ } %
\def\cotan{{\rm cotan \, }}

\Title{\vbox{\hbox{SPhT 92/138}\hbox{hep-th/9211138}}}
{{\vbox {
%\centerline{}
\bigskip
\centerline{ Fusion potentials I} }}}

\bigskip

\centerline{P. Di Francesco} \bigskip \centerline{and} \bigskip
\centerline{J.-B. Zuber}\bigskip

\centerline{ \it Service de Physique Th\'eorique de Saclay
\footnote*{Laboratoire de la Direction des Sciences et
de la Mati\`ere du Commissariat \`a l'Energie Atomique.},}
\centerline{ \it F-91191 Gif sur Yvette Cedex, France}

\vskip .2in
\noindent  We reconsider the conjecture by Gepner that the fusion
ring of a rational conformal field theory is isomorphic to
a ring of polynomials in $n$ variables quotiented by an ideal
of constraints that derive from a potential. We show that in a
variety of cases, this is indeed true with {\it one-variable}
polynomials.

\Date{11/92, submitted to J.Phys. A}
%\draftmode
%
%References
\lref\DGe{D. Gepner, Comm. Math. Phys. {\bf 141} 381-411 (1991).}
\lref\EV{E. Verlinde, Nucl. Phys. {\bf B300} [FS22] 360-376 (1988). }
\lref\GMDV{R. Dijkgraaf and E. Verlinde, Nucl. Phys. (Proc. Suppl.)
{\bf 5B} 87-97 (1988)\semi G. Moore and N. Seiberg, {\it Lectures on RCFT},
in {\it Superstrings {\oldstyle 89}},
proceedings of the 1989 Trieste spring school, M. Green {\it et al.} eds,
World Scientific 1990,  and further references therein. }
\lref\Ka{T. Kawai, Phys. Lett. {\bf B217} 247-251 (1989).}
\lref\Rav{M. Caselle, G. Ponzano and F. Ravanini, Phys. Lett. {\bf B251}
260-265 (1990)\semi
 {\it Towards a classification of fusion rule algebras in rational
conformal field theories}, Saclay preprint SPhT/91-174, Torino DFTT 46/91,
Int.J.Mod.Phys. {\bf B6} 2075 (1992)
%{ (hep-th/9111027)} and further references therein,
and further references therein. }
\lref\Bou{M. Bourdeau, E.J. Mlawer, H. Riggs and H.J. Schnitzer, Mod. Phys.
Lett. {\bf A7} 698-700 (1992).}
\lref\GSc{D. Gepner and A. Schwimmer, Nucl. Phys. {\bf B380} 147-167 (1992).}
\lref\Cr{M. Crescimanno,
{\it Fusion potentials for $G_k$ and handle squashing},
MIT preprint CTP\# 2021. %Nucl. Phys. {\bf B} ...?
}
\lref\Intr{K. Intriligator, Mod. Phys. Lett. {\bf A6} 3543-3556 (1991).}
\lref\AZ{A.B. Zamolodchikov, Sov. J. Nucl. Phys. {\bf 44} 529-533 (1986)
[Yad.Fiz. {\bf 44} 821-827 (1986)]. }
\lref\Gin{P. Ginsparg in
{\it Fields, Strings and Critical Phenomena}, Les Houches 1988,
E. Br\'ezin and J. Zinn-Justin eds., North-Holland 1990.}

\newsec{Introduction}

\noindent
The fusion properties constitute an essential piece of information on a
rational conformal field theory (RCFT). A few years ago, E. Verlinde \EV\ was
able to express the fusion rules in terms of the unitary matrix $S$
that encodes the modular transformations of the characters of the RCFT
\eqn\Ia{ N_{ij}^{\ k}=\sum_l {S_{il}S_{jl}S_{kl}^*\over S_{1l}}}
where ``1'' refers to the identity operator, and the labels $i,\cdots,l$ run
over $n$ values corresponding to the primary fields of the (extended) chiral
algebra of the RCFT \GMDV. The fusion coefficients
$N_{ij}^k $ are the structure constants of a commutative and associative
algebra
\eqn\Ib{ A_i A_j =\sum_k N_{ij}^{\ k} A_k \ .}
The matrices $N_i$ defined
by
\eqn\Ic{ \(N_i\)_{jk}= N_{ij}^{\ k}}
form themselves a representation of the fusion algebra
\eqn\Id{N_i N_j =\sum_k N_{ij}^{\ k} N_k}
as follows from the unitarity of the matrix $S$; this expresses the
associativity property of the algebra \Ib.
Relation \Ia\ implies that the matrix $S$ diagonalizes the matrices
$N_i$ and that their eigenvalues are of the form
\eqn\Ie{\Gc_i^{(l)}={S_{il}\over S_{1l}} \ .}

The general study of these fusion algebras \Ka\ and their classification
have been the object of much work \Rav. In particular, the possibility that
they may be represented by sets of polynomials has been considered \Rav.
In fact, a fusion algebra is a very special associative and commutative
algebra, as it possesses a selected basis corresponding to the primary fields
of the theory, in which the structure constants
are {\it non negative integers}.
For these reasons, one is not willing to trade this basis for
another one (except in those cases where there is a symmetry between
several primary fields).
Also one wants to stress the addition and
internal multiplication of fields rather than their multiplication by
scalars, {\it i.e.} the structure of {\bf ring} rather than that of
{\bf algebra}.

Recently, Gepner has made the conjecture that in any RCFT, the fusion ring
is isomorphic to a ring of polynomials in $p$ variables quotiented
by an ideal of constraints that derive from a potential \DGe.
He was able to prove this for the theories with a $SU(N)$ current algebra,
for which it is natural to take the $p=N-1$ variables associated to the
fundamental \rep s. This has been then extended to several other cases
\GSc\Cr\Bou\Intr.

Particularly interesting are faithful \rep s which have a trivial kernel,
that is in which no linear combination of the generators is represented
by the zero polynomial.
In the present note we discuss the possibility that the fusion ring
% polynomial ring be generated by a
be faithfully represented by a polynomial ring in a
single variable that corresponds to one of the primary fields of the theory
and is subject to a polynomial constraint that
may of course be integrated to a potential. We shall first give a simple
necessary and sufficient condition for this to happen, namely that
the eigenvalues of one of the fusion matrices be non degenerate. Curiously,
cases such as the models with a $SU(3)$ current algebra that
are known to have a fusion ring described by polynomials in {\it two}
variables, turn out to satisfy this condition and may thus be also
described by polynomials in a single variable. Even when the previous
condition is not fulfilled, in many cases, there is a way out, that
still enables one to use a single variable, at the expense of defining more
carefully what is meant by linear combination in the above definition of a
faithful \rep. This will be illustrated
on the minimal ($N=0$) models, and on the $D$ series of $SU(2)$ models
($\IZ_2$ orbifolds).

%%%%%%%%%%%%%%%%%%%%%%%%%%%%%%%%%%%%%%%%%%%%%%%%%%%%%%%%%%%%%%%%%%%%%%%%%%%%%
\newsec{A necessary and
 sufficient condition for a one-variable polynomial ring }

\noindent There exists a class of RCFT's in which the existence of a \rep\
of the fusion algebra by polynomials in a single variable may be ascertained.
Assume that among the matrices $N_i$, $i=1,\cdots, n$, there exists at least
one, call it $N_f$, with non degenerate eigenvalues. In other words,
the numbers $\Gc_f^{(l)}$ are all distinct.
%As all matrices $N_i$ commute, they
%may be diagonalized in the same basis (in fact, provided by the $S$'s), and
%for any $
Any other $N_i$ may be diagonalized in the same basis as $N_f$ and
there exists a unique polynomial $P_i(x)$ of degree at most $n-1$
such that its eigenvalues
$\Gc_i^{(l)}$ read
\eqn\IIa{\Gc_i^{(l)} = P_i\(\Gc_f^{(l)}\)}
$P_i$ being given by the Lagrange interpolation formula. Therefore, any
$N_i$ may be written as
\eqn\IIb{N_i = P_i(N_f)}
with a polynomial $P_i$; as both $N_i$ and $N_f$ have integral entries,
$P_i(x)$ must have rational coefficients. % in $\IQ$.
% solve the linear system for the coefficients !

The $n\times n$ matrix $N_f$, on the other hand, satisfies its
characteristic equation $\CP(x)=0$, that is also its minimal equation,
as $N_f$ has no degenerate eigenvalues. The constraint on $N_f$ is thus
\eqn\IIc{\CP(N_f)=0 }
that may of course be integrated to yield a ``potential'' $\CV(x)$,
\eqn\IId{\CV'(x)= \CP(x)\ ,}
that is a polynomial of degree $n+1$.

\vskip12mm
This situation is far from exceptional. We illustrate it with a few
examples~: \par
\noindent
* Ising model. Take for $f$ the spin $\Gs$ field. The fields $\un$, $\Gs$
and $\Ge$ obey fusion rules
\eqn\IIe{\Gs.\Gs=\un+\Ge \qquad \Gs.\Ge= \Gs \qquad \Ge.\Ge=\un}
that may be represented by the polynomials in the variable $\Gs$
\eqn\IIf{1 \qquad \Gs \qquad \Ge=\Gs^2-1}
subject to the constraint
\eqn\IIg{\Gs^3=2 \Gs}
that is derived from the potential
\eqn\IIh{\CV(\Gs)= {\Gs^4\over 4} -\Gs^2 \ . }

\noindent
* $SU(2)_k$~: this case is well known. (We are considering here the
``diagonal'' case, labelled by the $A_{k+1}$ Dynkin diagram.)
At level $k$, there are $k+1$ \rep s labelled by the integer
$n=2j$, $0\le n\le k$ and the fusion rules are represented by the
multiplication of \Che\ polynomials of the second kind, $P_n(x)$~:
\eqn\chebi{ P_n(2\cos \Gth) = {\sin(n+1)\Gth \over \sin \Gth}\ , \quad 0\le
n\le k}
with the constraint that
\eqn\constr{P_{k+1}(x)=0\ .}
\par

\noindent
* $SU(3)_k$~: We take for $f$ one of the two fundamental  \rep s.
The eigenvalues are known to be of the form
\eqn\IIi{\Gc_f^{(\Gl)}=\exp i\Ga_1 +\exp i\Ga_2 +\exp-i(\Ga_1+\Ga_2)\ , }
where the angles $\Ga_1$ and $\Ga_2$ read
\eqn\consang{\Ga_1=2\pi{\Gl_1-\Gl_2\over
3(k+3)}\ \  {\rm and} \ \ \Ga_2=2\pi{\Gl_1+2\Gl_2\over 3(k+3)},
\Gl_1,\Gl_2 \ge 1, \
\Gl_1+\Gl_2 \le k+2.}
It is a tedious but straightforward exercise of trigonometry to check
that two numbers of the form \IIi\ cannot be degenerate %\foot{
(see appendix A for the detailed proof). %}.
(In contrast, for $SU(4)_2$, it is easy to see that 0 is doubly degenerate
if we choose again for $f$ the fundamental \rep.)  We thus conclude
that the fusion ring of $SU(3)$ current algebras may be represented by
one-variable polynomials. As it is also naturally represented by two-variable
polynomials \DGe, (the two variables being associated with the two fundamental
\rep s), we have here a first instance of the non uniqueness of this
description of fusion rings by polynomial rings. For the sake of
definiteness, we give the expressions of the various fields
% FOOTNOTE
\foot{We shall henceforth make a slight abuse of notation and denote
in the same way the fields $\Phi$ regarded as generators of the fusion
algebra and their representatives.}\
 of $SU(3)_2$
labelled by their indices $\Gl_1, \Gl_2\ge 1$, $\Gl_1+\Gl_2\le 4$
\eqn\IIj{
\eqalign{\Phi_{(1,1)}&=1\qquad \Phi_{(2,1)}=x\qquad \Phi_{(3,1)}=\oh(5x^2
-x^5)\cr \Phi_{(1,2)}&=\oh(x^5-3x^2) \qquad \Phi_{(2,2)}=\oh(x^3-1)
\cr
\Phi_{(1,3)}&=\oh(x^4-3x) %{5\over 2}x^4-\oh x^7-x
\ ,\cr}}
with the constraint
\eqn\IIk{x^6-4x^3-1=0 }
arising from the potential $V={x^7\over 7}-x^4-x$.
Note that the $\IZ_3$ charge (the ``triality'') of each field may be read
off the degrees modulo 3 in the monomials of its expression in terms of $x$.

Although the number of variables and the form of the potential differ
from that given by Gepner, the ultimate content of the constraint, namely
the absence of certain \rep s in the Kac-Moody algebra, is the same and
the constraint and the polynomial \rep\ of the fields match after
elimination of one variable in Gepner's formulation.
Keeping the case of $SU(3)_2$ for illustration, we recall that the
two-variable potential of Gepner reads
\eqn\IIl{V(x,y)= {x^5\over 5} -x^3 y +x^2+x y^2-y}
from which follow the constraints
\eqna\IIm
$$\eqalignno{x^4-3x^2y+2x+y^2&=0 &\IIm a\cr
x^3-2xy+1&=0 \ & \IIm b\cr}$$
that express the absence of \rep s corresponding to Young tableaux
with more than two columns. The fields are represented as
\eqn\IIn{
\eqalign{\Phi_{(1,1)}&=1\qquad \Phi_{(2,1)}=x\qquad \Phi_{(3,1)}=x^2-y
\cr \Phi_{(1,2)}&=y \qquad \Phi_{(2,2)}=xy-1
\cr
\Phi_{(1,3)}&=y^2-x %{5\over 2}x^4-\oh x^7-x
\ .\cr}}
Eliminating $y$ between the two equations \IIm{} we find the equation
\IIk\ in $x$. The latter may be written as $x(x^5-4x^2)=1$, implying that
$x$ may be inverted on the ring. This makes it possible to solve in
$y$ using \IIm{b} and one recovers all the expressions \IIj.

\vskip13mm
Conversely let us suppose that one of the $N$'s, call it $N_f$,
generates the $n$ matrices $N_i=P_i(N_f)$ as linearly independent
polynomials on $\IC$. If $N_f$ had some degenerate eigenvalue, then
its minimal polynomial $\CM$ would be of degree at most $n-1$,
which means that one could construct at most $n-1$ linearly independent
polynomials of $N_f$, in contradiction with the above.
Therefore the condition that some $N$ has non degenerate eigenvalues
is necessary and sufficient to ensure that the fusion ring is
faithfully represented by one variable polynomials.

One could object that even if all $N$'s have degenerate eigenvalues,
one could still take some linear combinations thereof having all
eigenvalues distinct. This possibility will be discarded in the
present study in which we insist that the variable of the polynomials
corresponds to one of the primary fields of the theory; in other words,
the polynomial $P_1(x)=x$ is a representative of one of the fields. %??

We have stressed in the introduction the importance of the selected
basis of the fusion ring. This explains why situations described
by potentials of the same degree, for instance
$SU(3)_2$ of eqn. \IIj, with potential \IIk\
$V(x)={x^7\over 7}-x^4-x$, and $SU(2)_5$ with potential \constr\
$W(x)={x^7 \over 7}-x^5+2x^3-x$, and thus ``local rings''
of the same dimension are regarded as inequivalent. In
fact, the potential is not a sufficient
information to reconstruct the fusion ring with its special basis and its
integral structure constants. %(toward the ``perturbed ring''?)

\vskip13mm
Let us examine now what happens when the necessary and sufficient
condition is not satisfied. As already mentioned, %we already saw above,
one cannot then represent all fields by
one-variable polynomials linearly independent over $\IC$.
It turns out that in many cases, one may represent them by polynomials
with coefficients in
$\bar{\IQ}$, some extension of $\IQ$ by some algebraic number(s),
the polynomials being linearly
independent over $\IQ$. This seems sensible since the operations in
the fusion ring imply only {\it integral} combinations of the fields.
We shall now present two families of RCFT's for which this happens~:
the minimal models and the non diagonal ``$D$'' $SU(2)$ models.

%%%%%%%%%%%%%%%%%%%%%%%%%%%%%%%%%%%%%%%%%%%%%%%%%%%%%%%%%%%%%%%%%%%%%%%%%%%%%
\newsec{Minimal models}

\noindent
In the generic $(p,q)$ minimal model (that is, with a central charge
$c=1-6(p-q)^2/pq$), the condition of last section is not satisfied
and the fusion matrices $N_i$ have degenerate eigenvalues. We
shall now see that fusion rules can nevertheless be faithfully represented
by one-variable polynomials with coefficients in some
algebraic extension of $\IQ$.

Let us consider the most general minimal theory $(2p+1,q)$ (one of the
integers has to be odd as they are coprimes).
Its primary fields $\Phi_{(r,s)}$ are labelled by integer Kac
indices $1\leq r \leq 2p$, $1\leq s \leq q-1$, and satisfy the reflection
symmetry
property: $\Phi_{(2p+1-r,q-s)}=\Phi_{(r,s)}$, leaving only $p(q-1)$
independent primaries.
One can generate the fusion of these fields in a very simple way:
\item{(i)} impose respectively the $SU(2)_{2p-1}$ and $SU(2)_{q-2}$ fusion
rules on the
fields $\Phi_{(r,1)}$, $1 \leq r \leq 2p$, and $\Phi_{(1,s)}$, $1 \leq s
\leq q-1$
\item{(ii)} deduce the fusions of the fields $\Phi_{(r,s)}=\Phi_{(r,1)} \times
\Phi_{(1,s)}$ by imposing the reflection symmetry property.

\noindent
{}From this procedure it is clear that all the fields in the Kac table can
be realized as
polynomials of the two basic fields $A\equiv \Phi_{(1,2)}$ and $B\equiv
\Phi_{(2,1)}$
%(here we use for $A$ and $B$ any representation of the fusion rules,
%for instance one  can think of $A$ and
%$B$ as two of the matrices $N_i$ introduced in the previous section)
provided the above reflection symmetry is satisfied.
More precisely, if we use the \Che polynomials $P_n$ of the second kind
defined in \chebi\
%
%\eqn\chebi{ P_n(2\cos x) = {\sin(n+1)x \over \sin x} }
%
then we have:
\eqn\poly{
\eqalign{
\Phi_{(1,s)} &= P_{s-1}(A) \cr
\Phi_{(r,1)} &= P_{r-1}(B) \cr
\Phi_{(r,s)} &= P_{s-1}(A) P_{r-1}(B) \cr }}
and the minimal equations for $A$ and $B$ read
\eqn\minab{P_{q-1}(A)=P_{2p}(B)=0 \ . }
{}To get the correct fusion rules,
we have to impose the reflection symmetry property
\eqn\refl{ P_{s-1}(A) P_{r-1}(B) = P_{q-1-s}(A) P_{2p-r}(B) \ .}
In particular, for $s=q-1$ and $r=1$, we must have
\eqn\consab{ P_{q-2}(A)=P_{2p-1}(B)\ .}
It is possible to prove that \minab\-\consab\ are sufficient to ensure
the reflection symmetry \refl. At this stage, we have obtained a
perfectly satisfactory \rep\ of the fusion ring by polynomials in
two variables subject to the constraints 	\minab-\consab.

In a search of \rep s  by one-variable polynomials, we
look for a solution of \consab\ in the form
\eqn\ansatz{ B= \Ga P_{q-2}(A) }
where $\Ga$ is a constant to be determined. Thanks to the $SU(2)_{q-2}$
fusion rules, we have $P_{q-2}(A)^2=P_0(A)=1$, therefore our ansatz for
$B$ satisfies $B^2=\Ga^2 P_0(A)=\Ga^2$.
The \Che polynomials of even (resp. odd) order are even (resp. odd),
therefore the minimal equation for $B$ implies $P_{2p}(\Ga)=0$.
Multiplying both sides of \consab\  by $B$, and using the $SU(2)_{q-2}$
and $SU(2)_{2p-1}$ fusion rules, we get:
\eqn\calc{ B P_{2p-1}(B)=P_{2p-2}(B)=P_{2p-2}(\Ga)=\Ga P_{q-2}(A)^2=\Ga}
so that $P_{2p-2}(\Ga)=P_1(\Ga)$. Recursively, we get $P_{2p-1-m}(\Ga)=
P_{m}(\Ga)$, $m=0,1,..,p$, which can all be obtained from the single equation
\eqn\defalph{ (P_p-P_{p-1})(\Ga)=0\ .}
Pick any solution of \defalph, then the reflection symmetry property is
ensured.
Namely for $r$ odd
\eqn\rodd{
\eqalign{
P_{2p-1-r}(B)&=P_{2p-1-r}(\Ga)=P_{r}(\Ga) \cr
P_r(B)&={P_r(\Ga) \over \Ga}B=P_r(\Ga)P_{q-2}(A) \cr
P_{2p-1-r}(B)P_{q-2-s}(A)&=P_r(\Ga)P_{q-2-s}(A) \cr
&=P_r(\Ga) P_{q-2}(A) P_s(A)=
P_{r}(B) P_s(A) \cr}}
and for $r$ even
\eqn\reven{
\eqalign{
P_{2p-1-r}(B)&={P_{2p-1-r}(\Ga) \over \Ga}B=P_r(\Ga)P_{q-2}(A)\cr
P_r(B)&=P_r(\Ga)\cr
P_{2p-1-r}(B)P_{q-2-s}(A)&=P_r(\Ga)P_{q-2}(A) P_{q-2-s}(A)
=P_r(\Ga)P_s(A)\cr  &=P_r(B)P_s(A)\ .\cr}}
We also have to choose the solution $\Ga$ of \defalph\ in order for
the fields to be linearly independent over $\IQ$.
Using the defining relation \chebi, \defalph\ is easily solved, and
we choose
\eqn\alch{ \Ga=2 \cos ({\pi \over 2p+1})\ .}
The Kac table can be described in the following way: the first column
$\Phi_{(1,s)}$ consists of $(1,P_1(A),P_2(A),...,P_{q-2}(A))$ where
all fields are independent;
the second column of $\Ga(P_{q-2}(A),P_{q-3}(A),...,1)$, the third one
of $P_2(\Ga)(1,P_1(A),...,P_{q-2}(A))$, etc... Hence the
linear independence over $\IQ$ of $(1,P_1(\Ga),P_2(\Ga),...,P_{p-1}(\Ga))$
implies automatically the linear independence of all the fields in
the first $p$ columns of the Kac table,
i.e. of all the primaries in the theory.

We have thus completed the proof that the fusion rules of the minimal models
may be represented faithfully on a ring of polynomials in the variable
$A$. We may illustrate it on the case of the $(4,5)$ minimal model (the
tricritical Ising model). There are 6 fields that may be represented
according to the previous method (Beware! here the
roles of $r$ and $s$ have to be interchanged in the previous formulae)
\eqnn\Isitr
$$\eqalignno{
\un=\Phi_{(1,1)} &=1 \qquad \Phi_{(1,2)}=\Ga (A^2-1)\cr
\Phi_{(2,1)}&=A  \qquad \Phi_{(2,2)}=\Ga A &  \Isitr\cr
\Phi_{(3,1)}&=P_2(A)=A^2-1 \qquad \Phi_{(3,2)}=\Ga \cr
}$$
where $\Ga$ is a primitive root of $P_2(\Ga)-P_1(\Ga)=\Ga^2-\Ga-1=0$, i.e.
$\Ga=2\cos {l\pi\over 5}$, $l=1$ or $3$ and $A$ satisfies the constraint
\eqn\Isicon{P_3(A)=A^3-2A=0\ .}
%$$2 \cos \pi/5= (1+\sqrt{5})/2\qquad 2\cos2\pi/5=(-1+\sqrt{5})/2$$
%

In fact there is another \rep\ of the $(4,5)$ fusion ring where the variable
corresponds to the $(1,2)$ field
\eqnn\Isitri
$$\eqalignno{\un=\Phi_{(1,1)} &=1 \qquad \Phi_{(1,2)}=B\cr
\Phi_{(2,1)}&={1\over \sqrt{2}} (B-1)(B^2+B-1)
\qquad \Phi_{(2,2)}={1\over \sqrt{2}}(B^2+B-1) &  \Isitri\cr
\Phi_{(3,1)}&=P_3(B)=B(B^2-2) \qquad \Phi_{(3,2)}=P_2(B)=B^2-1 \cr }$$
subject to the constraint
\eqn\isic{P_4(B)=B^4-3B^2+1=0\ .}
The choice of the determination of the square root of 2 is arbitrary
and its change reflects the fact that the fusion algebra admits a $\IZ_2$
automorphism under which $\Phi_{(2,s)}$, $s=1,2$ are odd.

This solution is not unique~: there is another way of representing this
fusion algebra by polynomials of $B$, with the same expression for the
$\IZ_2$ even fields, but
%
%\eqnn\Isitrr
%$$\eqalignno{\un=\Phi_{(1,1)} &=1 \qquad \Phi_{(1,2)}=A\cr
$$\Phi_{(2,1)}={1\over \sqrt{10}} (B-3)(B^2+B-1)
\qquad \Phi_{(2,2)}={1\over \sqrt{10}}(1-2B)(B^2+B-1)\ .\eqno{  \Isitri'}%\cr
%\Phi_{(3,1)}&=P_2(B)=B^2-1 \qquad \Phi_{(3,2)}=P_3(B)=B(B^2-2)\cr }
$$
\vskip8mm
As another evidence of this non-uniqueness, we shall now establish that
one may also represent the fusion ring by polynomials in a variable
corresponding to the $\phi=
\Phi_{(2,2)}$ field, the ``order parameter''. %This will turn out to be
When the two integers labelling the minimal model are odd, this  follows from
the analysis carried out in sect. 2, see the end of the footnote below.
So let us concentrate on the case where the integer $q$
is even, in particular  the
unitary models where $2p+1$ and $q$ have to be consecutive
integers. %\ref\FQS{D. Friedan, Z. Qiu and S. Shenker, ... }
Then, it is well known (sect. 8.5 of \Gin)
that the fusion algebra admits a  $\IZ_2$ automorphism (this
is what we have just used in the tricritical Ising model) under which the
fields $\Phi_{(r,s)}$, $s$ even, are odd. These $N=p\({q\over 2}-1\)$
fields thus form a subset of the fusion ring.
Call $\phi=\Phi_{(2,2)}$ and form all its odd powers
$\phi^{2n-1}$, $1\le n\le N$, they are $\IZ_2$ odd, thus linear
combinations of the $N$ fields $\Phi_{(r,s)}$, $s$ even. This linear system
may be inverted
% footnote
\foot{That this linear system may be inverted can be
seen by proving that among the eigenvalues of $\phi=\Phi_{(2,2)}$,
of the form $\gamma^{(r,s)}=
4\cos {\pi r\over 2p+1}\cos{\pi s\over q}$, only  zero is degenerate
($p$ times); it follows that the minimal polynomial in $\phi$ is of degree
$p(q-1)-(p-1)=2N+1$, whereas the linear dependence of $\phi, \cdots,
\phi^{2N-1}$ would lead to a smaller degree. Degeneracies  other than zero
cannot occur for number theoretic reasons~: if
$\gamma^{(r,s)}=\gamma^{(r',s')}$, the ratio
$\cos{\pi r\over 2p+1}/\cos{\pi r'\over 2p+1}$ has to be a rational,
which can occur only in the trivial cases $r=r'$ or $r'=2p+1-r$. We
relegate the proof of these assertions to Appendix B.
The same argument
shows that for non-unitary $(p,q)$ minimal models with both $p$ and $q$
odd, or for the $(3,q)$ models,
 $\Phi_{(2,2)}$ has non degenerate eigenvalues and is thus a generator
of the fusion algebra.}
to yield polynomial expressions %with rational coefficients %
(in $\IQ[\phi]$) of all these
$\Phi$. In particular $\Phi_{(1,2)}=P(\phi)$, but we have seen above that
all the fields may be expressed polynomially in terms of $A=\Phi_{(1,2)}$.

For example, in the $(3,4)$ (Ising) model, the constraint on $\phi$
is of degree
3, integrating to a quartic potential, for the $(4,5)$ (tricritical Ising),
the $\phi$-potential is sextic. Up to this point, these degrees seem to
agree with those expected for the Landau-Ginsburg potentials by the
argument of Zamolodchikov \AZ. Unfortunately, this breaks down for
the higher minimal models. In general for the $(m,m+1)$ unitary minimal
model we expect a degree $2N+2=\Big[{m^2+1\over 2}\Big]-m+2$
whereas Zamolodchikov's potential is of degree $2m-2$.

Note that all the irrational
coefficients that have appeared, $\Ga$ in \alch, or $\sqrt{2}$ or $\sqrt{10}$
in \Isitri,\Isitri' belong to the field $\CU$ obtained by adjoining to $\IQ$
all the eigenvalues $\Gc_i^{(\Ga)}$
of all the $N$'s. This is not an accident, and we
shall now prove this fact in general. Suppose we have found a \rep\ of the
fusion algebra of a RCFT on a ring of polynomials $P_i(X)$ quotiented by the
constraint $V'(X)=0$. As before, we assume that one of the $P$'s, call
it $P_f$, is simply $P_f(X)=X$, and that the constraint is the minimal
polynomial of degree $p$
satisfied by $N_f$: in other words, the roots of $V'$ are all
distinct, $x_0^{(l)}$. Now the $P_i(x_0^{(l)})$ form a one-dimensional
\rep\ of the fusion algebra, thus are nothing else than the eigenvalues
$\Gc_i^{(l)}$. In particular $x_0^{(l)}=\Gc_f^{(l)}$.
If we regard this set of equations as a linear system
in the coefficients of the polynomial $P_i$
$$  P_i(x_0^{(l)})=\sum_{k=0}^{p-1} P_{ik} x_0^{(l)\, k}= \Gc_i^{(l)}$$
this system may be inverted, since its determinant is the Vandermonde
determinant of the distinct roots $x_0^{(l)}$, and gives for the coefficients
$P_{ik}$ values in the field $\CU$ generated by the $\Gc_i^{(l)}$.

It would seem possible to reduce further the one-variable polynomial
\rep\ by substituting for the variable $X$ one of the roots $x_0^{(l)}$
of $V'(X)$. % of degree $p$.
This would lead to one-dimensional \rep s of the fusion
ring. If the polynomial $V'$ of degree $p$ is reducible on $\IQ$, however, the
choice of a root $x_0^{(l)}$ amounts to setting to zero a factor
of $V'$ of degree less than $p$, resulting in an unfaithful \rep.
In all the cases relative to minimal models considered above, the
constraint is indeed a reducible polynomial on $\IQ$: $P_{2k+1}(X)$
factors out $X$ whereas
$$ P_{2k}(X)= (-1)^k (P_k(X)-P_{k-1}(X))(P_k(-X)-P_{k-1}(-X)) \ .$$

%%%%%%%%%%%%%%%%%%%%%%%%%%%%%%%%%%%%%%%%%%%%%%%%%%%%%%%%%%%%%%%%%%%%%%%%%%%%%
\newsec{The $SU(2)$ models of $D$-type }
\noindent
For even level $k=2n$, the $SU(2)_k$ fusion admits a $\IZ_2$ automorphism
with the fields of half-integer spin ($2j$ odd) odd; the orbifold with
respect to this symmetry
gives rise to the so-called $D_{n+2}$ models \ref\jadis
{D. Gepner and E. Witten, Nucl. Phys. {\bf B278} 493-549 (1986).}
\ref\matuzalem{D. Gepner, Nucl. Phys. {\bf B287} 111-130 (1987)
\semi A. Cappelli, C. Itzykson and J.-B. Zuber, Nucl. Phys. {\bf B280}
[FS18] 445-465 (1987).}.
For $k=0\ \mod 4$, i.e. $n=2\nu$, these models have
block-diagonal genus-one partition functions, with the $\nu+2$ \rep s of their
{\it extended} algebra labelled by an integer $l$ between 0 and  $\nu$
%running over the exponents of the $D_{}$ Dynkin diagram \matuzalem\
($\nu$ twice degenerate). Their fusion algebra
is known to be described by the rules
\eqna\IVa
$$\eqalignno{\Phi_0 . \Phi_l&=\Phi_l & \IVa a\cr
\Phi_1. \Phi_l &= \Phi_{l-1}+\Phi_l+\Phi_{l+1}\qquad 1\le l\le \nu-2&\IVa b\cr
\Phi_1.\Phi_{\nu-1}&=\Phi_{\nu-2}+\Phi_{\nu-1}+\Phi_{\nu}^{(+)}
+\Phi_{\nu}^{(-)}& \IVa c\cr
\Phi_1.\Phi_{\nu}^{(\pm)}&=\Phi_{\nu-1}+\Phi_{\nu}^{(\mp)}& \IVa d\cr
\Phi_{\nu}^{(+)}.\Phi_{\nu}^{(-)}&=\Phi_{\nu-1}+\Phi_{\nu-3}+\cdots&\IVa e\cr
}$$
with  the other fusion rules
deduced from those by the associativity property.
 In particular, it follows from \IVa{b,d} that
\eqna\IVaa
$$\eqalignno{\Phi_{\nu}^{(+)}+\Phi_{\nu}^{(-)} &=
(\Phi_1-\Phi_0).\Phi_{\nu-1}-\Phi_{\nu-2} &\IVaa a \cr
(\Phi_1-\Phi_0).(\Phi_{\nu}^{(+)}+\Phi_{\nu}^{(-)}) &= 2\Phi_{\nu-1}\ .
& \IVaa b\cr}$$
{}From the orbifold picture, one expects this case to be related to the
$A_{2n+1}$ discussed in Sec. 2, and the even \Che polynomials to be
the appropriate objects. We set
\eqn\IVb{\CP_l(x)= P_{2l}(y) %\quad  0\le l\le \nu-1
 \qquad {\rm where} \quad x=y^2-1}
in such a way that
\eqn\IVc{\CP_0=1 \qquad \CP_1(x)=x\ .}
They satisfy the rules (cf. the tensor multiplication by spin 1 in $SU(2)$)
\eqn\IVe{ \CP_1 \CP_l = \CP_{l-1}+ \CP_l+\CP_{l+1}\ .}
%
%which realizes \IVa{b}.
{}To complete the identification with \IVa{}, we have  to define two
polynomials $\CP_{\nu}^{(\pm)}$ satisfying
$$\eqalignno{ \CP_{\nu}^{(+)} +\CP_{\nu}^{(-)}&= \CP_{\nu} &\IVa d'\cr
\CP_{\nu}^{(+)}.\CP_{\nu}^{(-)} &= \CP_{\nu-1}+\CP_{\nu-3}+\cdots \equiv
\CQ_{\nu}  &\IVa e' \cr} $$
and to impose on $x$ the constraint that follows from \IVaa{}
\eqnn\IVf
$$\eqalignno{\CR_{\nu} & \equiv (x-1) \CP_{\nu}- 2\CP_{\nu-1}\cr
&= \CP_{\nu+1}-\CP_{\nu-1}=0\ .&\IVf\cr}$$

{}We have just seen that $\CP_{\nu}^{(\pm)}$ have to be
solutions of the equation
\eqn\IVfa{ X^2- X \CP_{\nu} +\CQ_{\nu}=0 }
and we shall indeed now prove that this equation admits polynomial
solutions.
Using \IVe, it is easy to see that one may factor out $(1+x)$ in
the polynomial $\CR_{\nu}$
\eqnn\IVg
$$\eqalignno{\CR_{\nu}&=(1+x) Z_{\nu}& \IVg\cr Z_{\nu}&=\CP_{\nu}-2
\(\CP_{\nu-1}-\CP_{\nu-2}+\cdots +(-1)^{\nu-1}\CP_0\)\cr}$$
thus that $x^p Z_{\nu}=(-1)^p Z_{\nu}\ \mod\CR_{\nu}$ and that
\eqn\IVh{Z_{\nu}(x)^2=Z_{\nu}(-1) Z_{\nu}(x)\qquad \mod \CR_{\nu}\ . }
Now, $\CP_l(-1)=(-1)^l$ is established recursively, whence
$Z_{\nu}(-1)=(-1)^{\nu}(2\nu+1)$ from which we conclude that
\eqn\IVi{Z_{\nu}^2=(-1)^{\nu} (2\nu+1) Z_{\nu} \ .}
%
%{}From the expression
On the other hand, since the $\CP$'s are \Che polynomials (see \IVb), one
has
\eqn\IVj{\CP_{l}^2= \CP_0+\CP_1+\cdots+\CP_{2l}}
(``addition of two spins $l$''),
but from \IVf\ follows that $\CP_{\nu+i}=\CP_{\nu-i}$, $i=1, \cdots, \nu$,
whence
\eqn\IVk{\CP_{\nu}^2= \CP_{\nu}+2\(\CP_{\nu-1}+\CP_{\nu-2}+\cdots+\CP_0\)\ .}
We may thus assert from \IVi\ and \IVk\ that the discriminant of the equation
\IVfa\ is a perfect square
\eqnn\IVl
$$\eqalignno{\Delta&=\CP_{\nu}^2 -4\(\CP_{\nu-1}+\CP_{\nu-3}+\cdots\)\cr
&= \CP_{\nu}-2\(\CP_{\nu-1}-\CP_{\nu-2}+\cdots\) \cr
&= Z_{\nu}    & \IVl\cr
&={(-1)^{\nu}\over 2\nu+1} Z_{\nu}^2\cr }$$
and that the two desired polynomials  $\CP_{\nu}^{(\pm)}$  read
\eqn\IVm{\CP_{\nu}^{(\pm)}= \oh\big[ \CP_{\nu}\pm {i^{\nu}\over\sqrt{2\nu+1}}
Z_{\nu}\big]\ .}
We conclude that the polynomials
$$ \CP_l, \qquad 0\le l \le \nu -1;\quad \CP_{\nu}^{(\pm)} $$
form a \rep\ of the fusion algebra of the $D_{2\nu+2}$ models of
$SU(2)$ current algebra.
That the diagonalization of the $D_{2\nu+2}$ fusion rules involves imaginary
coefficients whenever $\nu$ is odd is a well known fact (see for instance
\ref\BYZ{R. Brustein, S. Yankielowicz and J.-B. Zuber, Nucl. Phys. {\bf B313}
321-347 (1989).}).
The last point that we want to make is that these polynomials are
not independent on $\IQ$ when $2\nu +1$ is a perfect square.
In that case the \rep\ of the fusion algebra by the $\CP$'s is not
faithful. The first instance occurs for $D_{10}$, for which
the combination of fields
$$ X=4\Phi_4^{(-)}-2\Phi_4^{(+)}-2\(\Phi_3-\Phi_2+\Phi_1-\Phi_0\)\ ,$$
that has the property that for any field $Y$, $X.Y=\lambda_Y X$,
($\lambda_Y\in \IZ$), is represented by 0.

For illustration we provide explicit formulae for the $D_4$ and $D_6$ cases;
the case of $D_4$ is a bit exceptional, as the generator of the ring is
one of the ``degenerate'' fields
\eqn\Dqu{D_4\ : \qquad \Phi_0=1,\ \Phi_1^{(+)}=X, \ \Phi_1^{(-)}=X^2
\qquad  X^3=1}
whereas $D_6$ exhibits the phenomena discussed above
\eqn\Dsix{D_6\ : \Phi_0= 1; \  \Phi_1=\phi; \ \Phi_2^{(\pm)}= \oh\Big[
(\phi^2-\phi-1)\pm {1\over \sqrt{5}} (\phi^2-3\phi+1)\Big]}
$$  {\rm with }\qquad \phi^3-2\phi^2-2\phi+1=0\ .$$

The cases of the $E_6$ and $E_8$ $SU(2)$ theories are readily
dealt with if one observes that their fusion algebras are
respectively isomorphic to those of the $(3,4)$ (Ising) and
$(2,5)$ (Lee-Yang) minimal models for which the discussion of the
previous section applies.

%$(A,D)$, $(A,E)$ minimal models?? (ras le bol!)

%%%%%%%%%%%%%%%%%%%%%%%%%%%%%%%%%%%%%%%%%%%%%%%%%%%%%%%%%%%%%%%%%%%%%%%%%%%%%
\newsec{Conclusion and questions}
%C'est bien joli mais \`a quoi \c ca sert tout \c ca???
\noindent
In this note we have shown that in several classes of rational conformal
field theories the fusion ring may be represented as a ring of polynomials
in one variable, quotiented by a certain polynomial constraint that may
be integrated to a ``potential''. This extends to cases like the $SU(3)$
models or the ``$D$'' series of $SU(2)$ models which were generally believed
to require two-variable polynomials. In most cases, the solution to Gepner
conjecture does not seem to be unique, and even the degree of the polynomial
constraint depends on the generator of the ring. In fact, this non-uniqueness
shatters the too
na\" \i ve idea that was one of the original motivations of the present work,
namely that this fusion potential might serve to {\it characterize} the
fusion ring.

In minimal models, in particular, we have seen that there are quite a
number of alternative ways of constructing this polynomial representation
and the corresponding potential.
Beside the information they encode on the fusion algebra, it would be quite
interesting to find a {\it physical} (Landau-Ginsburg?) interpretation
of these potentials.

Not all RCFT's admit such a \rep\ of their fusion ring~: {\it v.i.z.}
the $D_{10}$ orbifold whose \rep\ is not faithful, or $SU(4)_2$... It
would be nice to have a characterization of those RCFT's that have
this property.

In a forthcoming paper, we shall comment on some connections between this
fusion potentials and the potentials that emerge from perturbed topological
field theories.

%%%%%%%%%%%%%%%%%%%%%%%%%%%%%%%%%%%%%%%%%%%%%%%%%%%%%%%%%%%%%%%%%%%%%%%%%%%%%
\vskip12mm{\bf Acknowledgements.}
Special thanks are due to M. Bauer for his interest and generous assistance.
We are also very grateful to  C. De Concini, C. Itzykson and H. Kn\"orrer for
several judicious observations and suggestions.
\par
\appendix{A}{}

\noindent Given the rational numbers $\Ga_1$ and $\Ga_2$, let us look for
rational solutions $\Gb_1$ and $\Gb_2$ to%
\eqn\puzz{ e^{i\Ga_1}+e^{i\Ga_2}+e^{-i(\Ga_1+\Ga_2)}=
e^{i\Gb_1}+e^{i\Gb_2}+e^{-i(\Gb_1+\Gb_2)} }
Multiplying both sides of \puzz\ by
$e^{{i\over 2}(\Ga_1+\Ga_2+\Gb_1+\Gb_2)}$, and taking the real
and imaginary parts, we are left with the system
\eqn\syst{
\eqalign{
\sin\big({\Ga_1-\Gb_1 \over 2}\big)\sin\big(\Ga_1+\Gb_1+
{\Ga_2+\Gb_2 \over 2}\big)+\sin\big({\Ga_2-\Gb_2 \over 2}\big)
\sin&\big(\Ga_2+\Gb_2+{\Ga_1+\Gb_1 \over 2}\big)=0 \cr
\sin\big({\Ga_1-\Gb_1 \over 2}\big)\sin\big({\Ga_1+\Ga_2+\Gb_1 \over 2}\big)
\sin\big({\Ga_1+\Gb_1+\Gb_2 \over 2}\big)&+\cr
+\sin\big({\Ga_2-\Gb_2 \over 2}\big)
\sin\big({\Ga_1+\Ga_2+\Gb_2 \over 2}\big)
&\sin\big({\Gb_1+\Gb_2+\Ga_2 \over 2}\big)=0  \cr}}
Introduce the variables
\eqn\varia{
\eqalign{ a={\Ga_1+\Ga_2+\Gb_1 \over 2} \qquad &b={\Gb_1+\Gb_2+\Ga_1
\over 2} \cr
c={\Ga_1+\Ga_2+\Gb_2 \over 2} \qquad &d={\Gb_1+\Gb_2+\Ga_2 \over 2}\cr}}
then eqns. \syst\ read
\eqna\resys{
$$\eqalignno{
\sin(c-d)\sin(a+b)&= \sin(b-a)\sin(c+d) &\resys a\cr
\sin(c-d)\sin a \sin b &= \sin(b-a) \sin c \sin d &\resys b\cr}$$}
Note that the vanishing of any sine factor on each side of
\resys b\ (e.g. $c-d=b-a=0 \ \ \mod\pi$) just amounts to
taking one of the six trivial solutions of \puzz, namely (all the
equalities are understood modulo $\pi$)
$\Gb_1=\Ga_1, \ \Gb_2=\Ga_2$, or $\Gb_1=\Ga_1, \ \Gb_2=-(\Ga_1+\Ga_2)$
or the four permutations obtained by letting $\Ga_1 \leftrightarrow
\Ga_2$ or $\Ga_1 \leftrightarrow -(\Ga_1+\Ga_2)$.
Let us exclude these trivial solutions, then we can divide \resys a\
by \resys b\ to get
\eqn\divid{
{\sin(a+b) \over \sin a \sin b} = { \sin(c+d) \over \sin c \sin d} }
and rewrite \resys b\ as
\eqn\rewr{
{\sin(b-a) \over \sin a \sin b} = {\sin(c-d) \over \sin c \sin d} }
hence we get $\cotan a = \cotan d$ and $\cotan b= \cotan c$,
already excluded above.
In conclusion the only solutions to \puzz\ satisfy
$\{\Gb_1,\Gb_2,-(\Gb_1+\Gb_2)\}=\{\Ga_1,\Ga_2,-(\Ga_1+\Ga_2)\}$
and reduce to $\Gb_i=\Ga_i$ when the angles are constrained by
\consang.
This completes the proof that the $SU(3)_k$ fusion rules may be represented
by polynomials
in one variable.

\appendix{B}{}

\noindent For\foot{This proof is entirely due to Michel Bauer.}
two given coprime integers $p$ and $q$, and $r$ and $s$
such that $1 \leq r <p/2$,
$1 \leq s < q/2$, we look for solutions $r'$ and $s'$ to
\eqn\puzzle{ \cos \pi {r \over p} \cos \pi {s \over q}
= \cos \pi {r' \over p} \cos \pi {s' \over q} }
This amounts to
\eqn\amount{ {\cos \pi {r \over p} \over \cos \pi {r' \over p}}=
{ \cos \pi {s' \over q} \over \cos \pi {s \over q}} }
the value of \amount\ belongs to the intersection of the
extensions $\IQ(\cos {\pi \over p}) \cap \IQ(\cos {\pi \over q})$,
and is therefore rational.
Let us show that indeed $\IQ(e^{i \pi \over p}) \cap
\IQ(e^{i \pi \over q})=\IQ$.
Set $\xi=e^{i\pi \over pq}$, $\Ga=\xi^q$, $\Gb=\xi^p$.
Suppose we have
\eqn\supp{ \sum_{i=0}^{2p-1} a_i \Ga^i-\sum_{j=0}^{2q-1} b_j
\Gb^j = f(\xi) =0 } %_{\{a_i\},\{b_j\}}(\xi)=0}
where $a_i$ and $b_j$ are rational. The equation $f(\xi)=0$
is algebraic with rational coefficients,
and is therefore satisfied by any conjugate
$\xi^c$, $1 \leq c<2pq$, $c$ coprime with $2pq$.
In particular let us choose those which preserve $\alpha=\xi^q$,
namely take $c$ to be of the form $c_k=1+2kp$, $k$ defined modulo $q$.
It is straightforward to see that there exists a subset
$K$ of $\{0,1,..,q-1\}$, such that $\{c_k, \ k \in K\}$
runs over %describe exactly
all the integers coprime with $2q$ modulo $2q$,
which means that $\beta^{c_k}$ runs over %describe exactly
all the conjugates of $\beta$.
Thus if we write
\eqn\zer{ 0 = {1 \over |K|}\sum_{k \in K} f(\xi^{c_k}) =
\sum_{i=0}^{2p-1} a_i \Ga^i -{1 \over |K|}\sum_{k \in K}
\sum_{j=0}^{2q-1} b_j {\beta}^{jc_k}}
the second sum in the r.h.s. of \zer\ is a symmetric function of $\Gb$
and its conjugates, therefore rational.
%(recall that $\Gb$ and its conjugates
%solve an algebraic polynomial equation).
Hence both sums in \supp\ are rational and we proved the statement.

We are now left with the problem of finding solutions to
\eqn\auxi{ {\cos \pi {r \over p} \over \cos \pi {r' \over p}} \in \IQ}
where, say $p=2l+1$ is an odd integer (otherwise exchange the roles of
$p \leftrightarrow q$, $r \leftrightarrow s$ and $r' \leftrightarrow s'$).
Consider the polynomial
\eqn\popol{ \Pi(x) = \prod_{r,r'=1}^{p-1}
(2 x \cos \pi {r' \over p} - 2 \cos \pi {r \over p} )\ .}
This polynomial has integer coefficients (as polynomial
symmetric functions of $\{e^{i \pi a/p}, \ a=\pm 1,..,\pm (p-1)\}$).
Moreover it is monic, due to the well known identity
\eqn\well{ \prod_{r=1}^{p-1} 2 \cos \pi {r \over p} = (-1)^l}
for odd $p=2l+1$,  of degree $(p-1)^2$
and reciprocal {\it i.e.} $\Pi(x)=x^{(p-1)^2}
\Pi\({1\over x}\)$.
The  root of a monic polynomial with integer coefficients
is what is called an algebraic integer;
if it is rational, it
has to be an integer
(otherwise one could write it $a/b$, and one would have
an equation $(a/b)^n= {\rm integer}/b^{n-1}$, which is impossible).
But if $a$ is a root of $\Pi(x)$, then $1/a$ is also a root, therefore
the only rational roots of $\Pi(x)$ are $\pm 1$ and thus
the only ways of realizing \auxi\ are by taking
$\cos \pi {r \over p} = \pm \cos \pi {r' \over p}$, i.e. $r'=r$ or
$r'=p-r$, from which we deduce that $s'=s$ or $s'=q-s$.
This completes the proof that the only possibly degenerate eigenvalue for
$N_{(2,2)}$ in the $(p,q)$ minimal conformal theory is zero.

\listrefs
\end